\documentclass[12pt]{article}

\usepackage[colorinlistoftodos]{todonotes}
\usepackage{graphicx}
\usepackage{float}
\usepackage{caption}
\usepackage[utf8]{inputenc}
\usepackage{notoccite}
\usepackage[margin = 1in]{geometry}
\usepackage{parskip} 
\usepackage{version}
\usepackage{comment}
\usepackage{array}
\usepackage{amsmath}
\usepackage[colorinlistoftodos]{todonotes}
\usepackage[colorlinks=true, allcolors=blue]{hyperref}
\usepackage[T1]{fontenc}

\title{The temporal evolution of venture investment strategies in sector space\footnote{We are grateful to G. Dion, T. Lacroix and R. Taub for several helpful discussions and to F. Krzakala from Ecole Normale Sup\'erieure for a very helpful suggestion.}}
\author{Th\'eophile Carniel\textsuperscript{1,2,3}, Cl\'ement Gastaud\textsuperscript{1,3}, Jean-Michel Dalle\textsuperscript{$\dagger$,1,3}}

\begin{document}
\maketitle

\bigskip
\textbf{1} Agoranov, Paris, France

\textbf{2} PSL Research University, Paris, France

\textbf{3} Sorbonne Universit\'e, Paris, France

\textbf{$\dagger$} jean-michel.dalle@sorbonne-universite.fr (J.-M. D.)

\bigskip

\begin{abstract}
We analyze the sectoral dynamics of startup venture financing. Based on a dataset of 52000 start-ups and 110000 funding rounds in the United States from 2000 to 2017, and by applying both Principal Component Analysis (PCA) and Tensor Component Analysis (TCA) in sector space, we visualize and measure the evolution of the investment strategies of different classes of investors across sectors and over time. During the past decade, we observe a coherent evolution of early stage investments towards a lower-tech area in sector space, associated with a marked increase in the concentration of investments and with the emergence of a newer class of investors called accelerators. We provide evidence for a more recent shift of start-up venture financing away from the previous one.
\end{abstract}

\section{Introduction}
Venture financing allows startups to survive and grow until product development is finished and/or critical mass in terms of market share is reached, i.e. until they become profitable \cite{Hellmann2000}. Venture capitalists also provide entrepreneurs with relevant coaching and advice with regard to various aspects of startup founding, management and growth, as well as access to business contacts and opportunities \cite{Gifford1998,Gompers2001}. Choosing investors therefore plays a critical role in the success or failure of a startup, and previously successful investors are sought after by entrepreneurs and rise in prominence within startup ecosystems and investor communities (e.g.~\cite{Pratt2018}).

Investor types in startup ecosystems mirror the alphabet round system, with investors specialized in so-called Seed (typically, a few hundreds of thousands of dollars or euros), Series A (typically, from 1 to 5 million dollars or euros), Series B (typically, from 5 to 30 million dollars or euros), Series C (typically, several tens of millions of dollars or euros) and later D, E, F, etc. rounds. Seed and Series A are known as early-stage investments, while later rounds constitute growth or late-stage investments. Typically, late-stage venture-capital funds tend to invest large amounts of money --- that they themselves have raised from various sources such as banks, insurance companies and other institutions --- in the form then of Series B or later rounds, in startups that have already grown to a significant size and need money to pursue their development further, while so-called early-stage funds operate similarly but at Series A stage.

For their part, angel investors are individuals who invest their own money in limited amounts and who operate mostly at Seed stage. Still among early-stage investors, a further and more recent addition to entrepreneurial ecosystems is related to the emergence of accelerators \cite{Pauwels2016,Cohen2014}, a new sub-type of investors that operates at Seed stage and follows a specific model, selecting a group (a cohort) of start-ups at a very early stage of development and providing them with coaching and education on matters relevant to entrepreneurship for a short period of time (typically between 3 and 6 months) in exchange for a few percents of their equity.

In addition, and due to the fact that they co-invest in startups, either at the same funding round or at sequential rounds, it has long been recognized that investors are embedded in networks~\cite{Hochberg2007,Sorenson2001}. As a consequence, their investment decisions and strategies affect and are affected by the investment strategies of other investors in their network and ecosystem~\cite{Jin2015}.

In this context, and somewhat surprisingly, the actual interactions between the individual investment strategies of all investors have not really been the subject of direct empirical studies. Popular assessments about "herding" behaviors or about investment fads and fashions are widespread, sometimes supported by anecdotal evidence, but we still cannot observe, measure or evaluate how, and in what respect, the investment strategies of investors, and of each kind of investors, coordinate and evolve through time. Whereas public financial markets have, on their part, been heavily studied in this respect, and mostly due to the unavailability of comprehensive datasets, the complexity of the venture ecosystem has only limitedly been subject to similar scientific investigations up to now. To put it yet differently, although new ventures have been a corner topic of the entrepreneurial research literature for the past 20 years~\cite{Chandra2018}, and even though investments strategies considerably structure the dynamics of startup ecosystems, we are still mostly missing methods and tools that would allow us, and stakeholders, to observe and analyze directly the global and temporal evolution of investor strategies, most notably among sectors.

In this paper, we analyze a comprehensive dataset of the venture financing ecosystem in the United States for the period 2000-2017 that includes detailed information on startups, notably sectoral tags and financing rounds, and we present a simple but novel analysis framework based on applying both principal component analysis (PCA) and tensor component analysis (TCA) to investment strategies in sector space. Within this framework, we are able to observe and characterize the dynamics of the strategies of different categories of investors and, most notably, the recent evolutions of early-stage investment strategies within sector space.

\section{Methods}
\subsection{Dataset and data processing}
\label{subsec:dataprocessing}

The dataset used was extracted in July 2018 from \href{http://crunchbase.com}{Crunchbase}, a popular and open data source for scientists studying the startup ecosystem (see~\cite{Dalle2017} for a survey). Crunchbase includes detailed information on startups (founding date, amount of money raised, categories describing the sectors in which the start-up operates, funding rounds and investors, etc.). In order to focus on coherent phenomena, we selected only US-based startups that were still active (i.e. had not been closed), that had raised money at least once, and that had been founded after January 1st, 2000, which resulted in 51 841 US-based startups. 

Based on the sectoral tags provided by Crunchbase for each startups, we created a tree-like sectoral ontology with 28 first-order "parent" tags i.e. tags not contained in any other sectoral tag.
The determination of these 28 parent sectors was done manually by parsing through Crunchbase category groups and deleting, fusing or reordering those that were not sufficiently descriptive and independent from one another. In some instances we also edited the sectoral, category tags of the startups in order to get rid of redundant and/or non-descriptive occurrences.

We reconstructed the portfolios of all investors present in the dataset (29278 investors of all types) and, with this information, we created for each year and each investor, a table containing information on the investments of this investor for this given year i.e. a non-normalized 28-dimensional dataset, with each dimension corresponding to a parent secotral tag, and containing the number of rounds and the total funding amount invested in that sector by that investor during each given year. This table therefore gives each investor's sectoral investment strategy for each given year, approached here through how many investments this investor has made in all sectors. Information about the stage (seed capital, series A, series B, etc) of the investments was also retained throughout the whole process.

It should be noted that, in order to allow for comparisons between different sectors, we looked at the number of rounds invested in each parent tag. Compared to their number, the funding amounts of these funding rounds is very sector-dependent as some sectors are more capital-intensive than others, which would have made comparisons between investors less reliable. When a single startup had tags in several different parent tags, we divided the investment between the parent tags equally. In addition, all investments pertaining to the \textit{Health Care} parent tag were excluded from the analysis as this sector's profile was found to be very different from others, with slower dynamics, highly-specialized investors and large funding amounts.

\subsection{Investment barycenters in sector space}
\label{subsec:PCA_method}

    By identifiying and classifying start-ups using sectoral tags, we are able to give each investor's strategy a position in a 28-dimensional space where each dimension is associated to a parent sectoral tag. This simple projection in sectoral space simplifies data handling and also enables the use of various data analysis techniques in terms of visualization and analysis. By aggregating data on all investors, we also estimate the barycenter of all investors' strategies for each given year, as defined in equation~\ref{eq:barycenter}.

\begin{equation}
    \label{eq:barycenter}
    X_{k} = \frac{1}{N}\sum_{i = investors}x_{i,k}n_{i} ,
\end{equation}

\small{where $X_{k}$ is the position of the barycenter in dimension k of the tag-space, $n_{i}$ the total number of rounds investor $i$ was part of for a given year and $N$ the total number of rounds by all investors in the given year.}

The evolution of these barycenters (centers of gravity), being intrinsically geometrical objects, is then easily studied through graphical representations and through techniques that allow for their mathematical manipulation. In order to do so, we first use Principal Component Analysis (PCA), a common dimensionality-reduction technique that takes into account correlations between dimensions. By finding the directions (i.e. the vectors) of maximal variance in sector space, it is indeed possible to reduce the dimensionality of our dataset by creating linear combinations of the existing directions that retain as much of the variance as possible and that are orthogonal to one another based on the correlations between the initial dimensions. Using these linear combinations and the associated set of coordinates in sector space, we project our dataset in a subspace of lower dimension and are able to visualize sectoral information related to investors' investment strategies and portfolio in the 2-D space obtained through the first two PCA orthogonal dimensions.
Specifically here, we created an array with the 28 parent tags as columns and normalized yearly distributions for all investors as lines. We standardized all columns to 0 mean and 1 standard deviation, as is common practice in PCA techniques~\cite{Standardization}.

\subsection{Analyzing temporal evolution with TCA}
\label{subsec:TCA_method}
The investment ecosystem as a whole is also an adaptive system, where trends come and go, some of which have a durable impact on the structure of the ecosystem. New actors and new strategies become part of it while others get left by the wayside if their outcomes are below expectations. Typically, new types of actors can be created (e.g. accelerators) or existing structures can see their functions change (e.g. a shift of strategies of institutional VCs) as a reaction to the environment of the system, for instance exogenous events such as the financial crisis of the late 2000s. During recent years and following notably~\cite{Williams2018}, TCA (Tensor Component Analysis) or MPCA (Multi-Dimensional Principal Component Analysis) have seen a gain of interest, notably applied to neural dynamics, in order to take into account such adaptation effects of a group of actors over a series of temporally-ordered measurements. In a similar way to PCA, TCA reduces a high-dimensional data \textit{tensor} into a lower-dimensional number of components $R$. Each of these components has 3 associated factors :

\begin{itemize}
    \item the \textit{investor factor}, that gives the weight of each individual investor in component $r \in R$
    \item the \textit{sector factor}, equivalent to PCA loadings on investment sectors
    \item the \textit{temporal factor}, that corresponds to the variation over time of the amplitude of activity patterns in relation to component $r$
\end{itemize}

\par
Following~\cite{Williams2018}, our dataset was restructured into a data tensor of dimension $N \times S \times K$, where the first dimension represents individual investors ($N = 27694$ unique investors for the United States between 2000 and 2017), the second dimension represents sectors ($S = 28$ sectors as described in section \ref{subsec:dataprocessing}) and the third dimension represents investor activity for all $K = 18$ years between 2000 and 2017 (included). To use a neuronal analogy, an individual investor \"spikes\" in the given year with activity profile $\textbf{n}_{k}$ corresponding to its investments, each year $k$ being considered as a new trial for investor $n$ with a potentially different investor profile $\textbf{n}_{k}$. Again, the usual standardization procedure was applied to each yearly matrix $\textbf{A}_{k}$, setting each feature's mean and standard deviation to 0 and 1, respectively.
\par
\begin{figure}
    \centering
    \includegraphics[scale = .25]{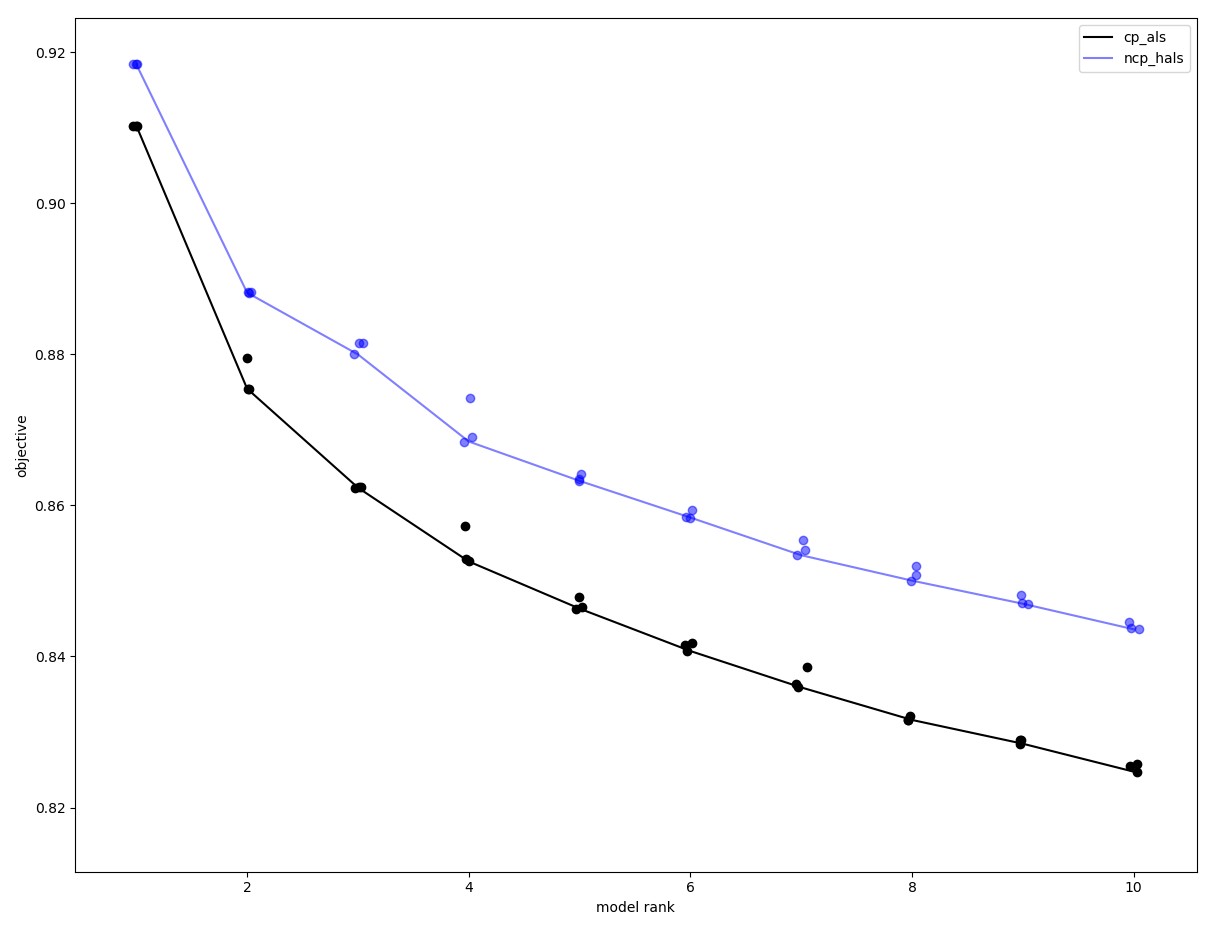}
    \includegraphics[scale = .25]{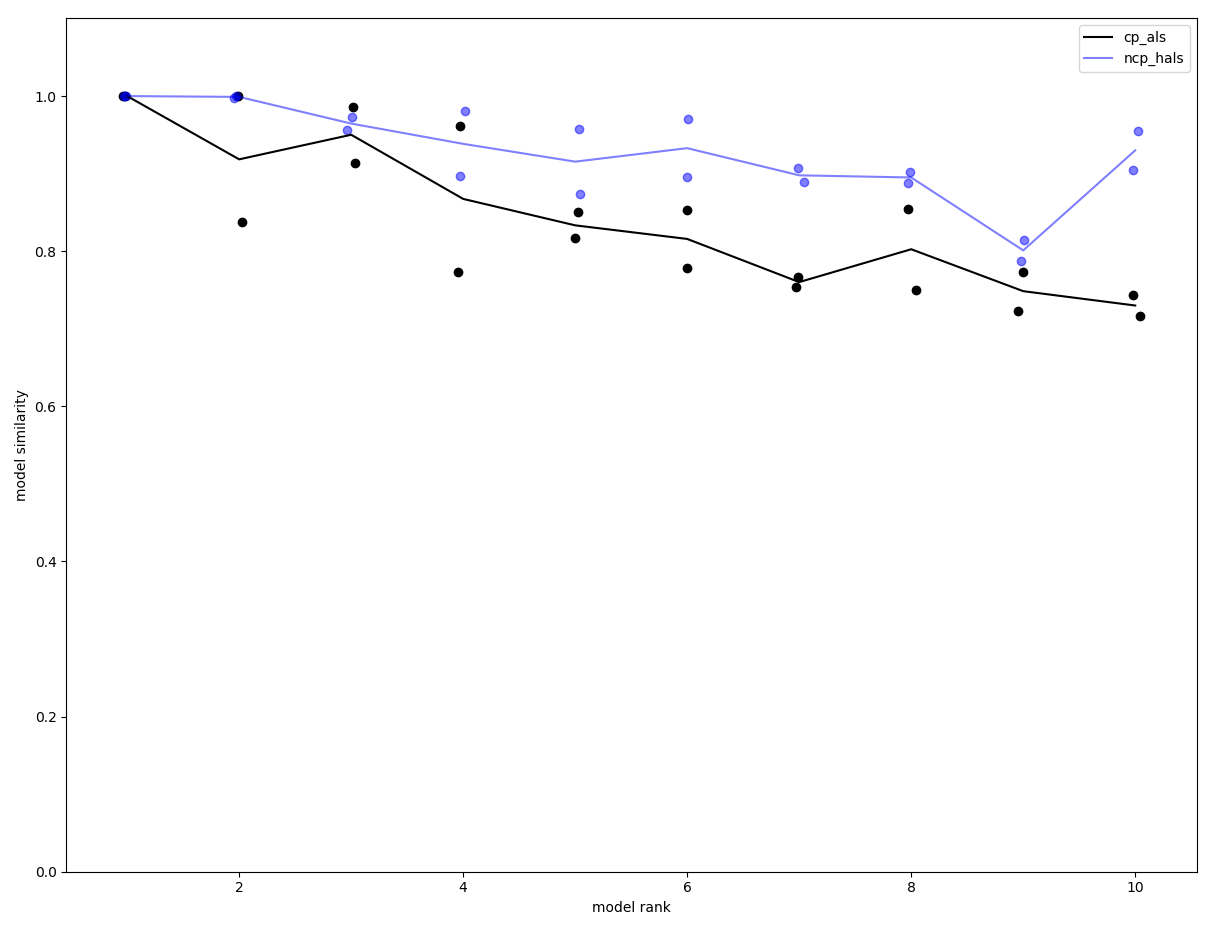}
    \caption{\small{Reconstruction error (left) and model similarity (right) as a function of number of components $R$.}}
    \label{fig:obj_sim_plots}
\end{figure}

Results from fitting our data with a tensor decomposition model are presented in fig.~\ref{fig:obj_sim_plots}. In order to determine the number of dimensions $R$ to be kept in the model, we selected the value of $R$ that maximizes both model similarity and the absolute value of the first-order derivative of the reconstruction error. Adding more components (increasing $R$) continually decreases reconstruction error, while minimizing model similarity implies low values of $R$. Looking for the maximum of the first-order derivative provides a value at which model similarity remains high while the accuracy gained from adding dimensions to our model starts experiencing diminishing returns.

\section{Results}
\label{sec:results}

\subsection{Temporal dynamics using PCA}
\label{subsec:results_PCA}

\begin{figure}[H]
    \raggedleft
    \includegraphics[scale = .9]{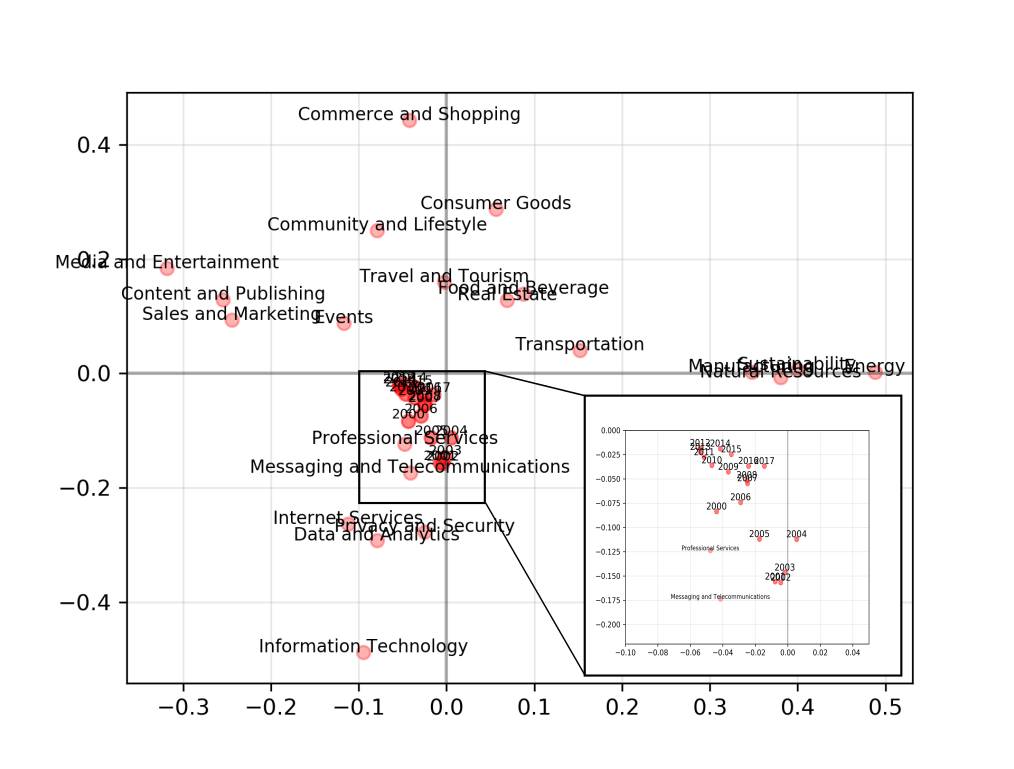}
    \caption{\small{Positions of sectoral tags in 2-D sectoral space and of the projected barycenters.}}
    \label{fig:pca_tags}
\end{figure}

Looking at the evolution of the projected position of the barycenter in 2-D sectoral space (fig.~\ref{fig:pca_tags}), we observe a shift towards the top-left quadrant, from the early 2000s until today. By positioning sectoral tags in 2-D sectoral space (fig.~\ref{fig:pca_tags}), we analyze this evolution as  corresponding to a displacement of the center of gravity of investment strategies towards more consumer-oriented ("B2C"), low-tech investment strategies as are commonplace in sectors such as \emph{Messaging \& Telecommunications} or \emph{Content \& Publishing}, away from more "deeptech" investments as they characterize both \emph{Energy} or \emph{Manufacturing} (as opposed to \emph{Sales \& Marketing} or \emph{Media \& Entertainment} on the x-axis of fig.~\ref{fig:pca_tags}), and \emph{Data \& Analytics} or \emph{Privacy \& Security} (as opposed to \emph{Commerce \& Shopping} on the y-axis of fig.~\ref{fig:pca_tags}).

\begin{figure}[H]
    \centering
    \includegraphics[scale = .5]{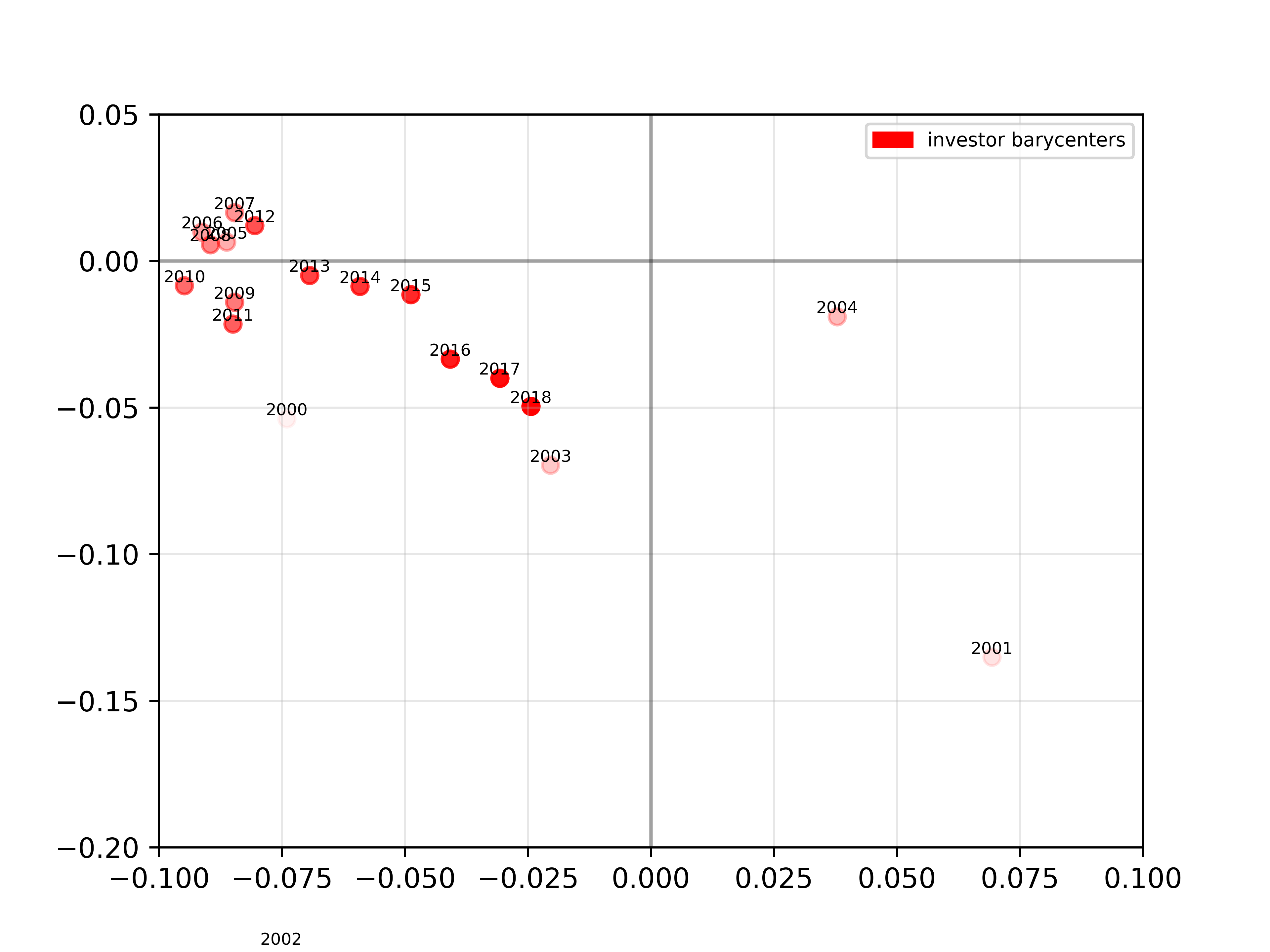}
    \includegraphics[scale = .5]{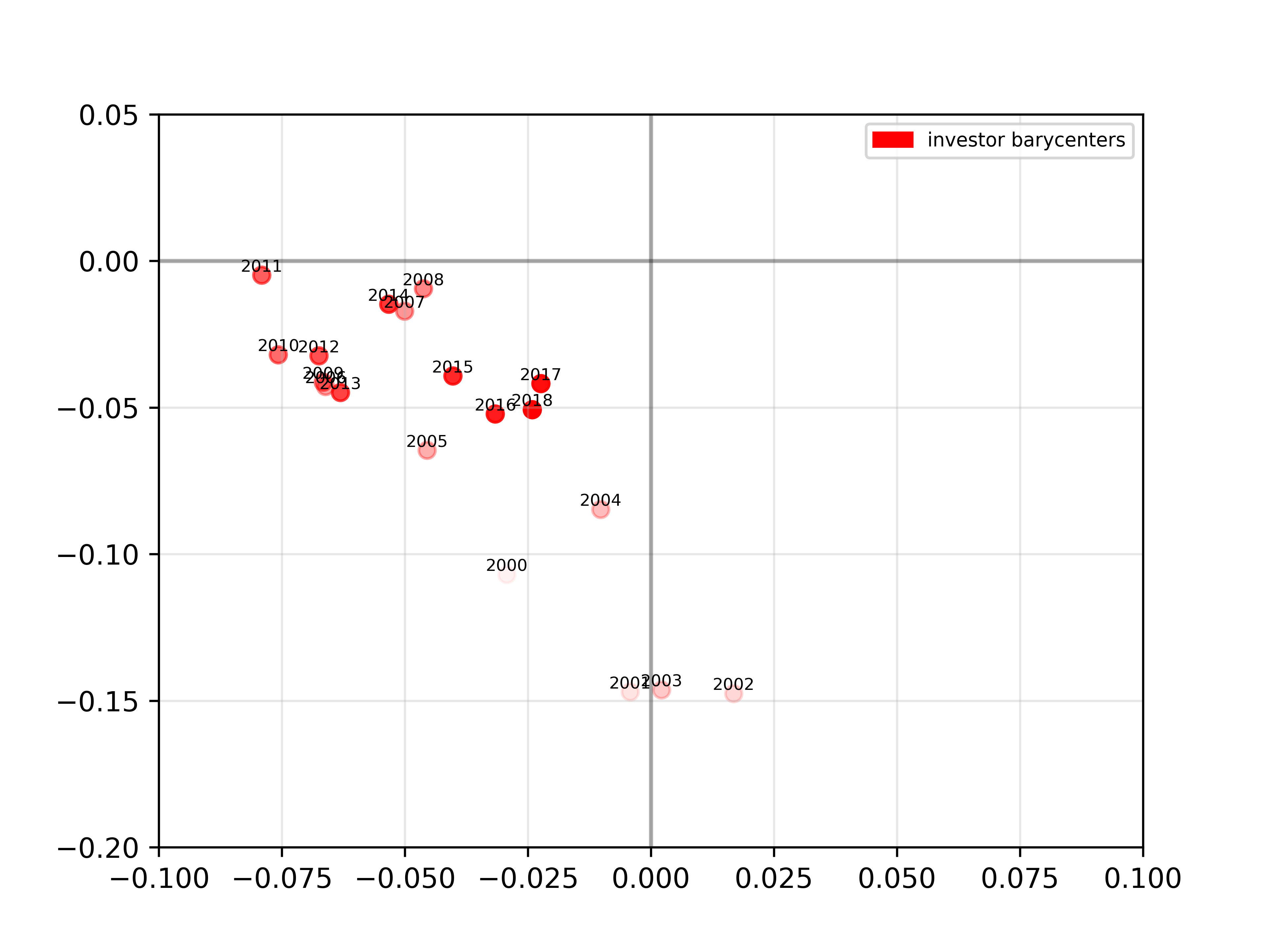}
    \includegraphics[scale = .5]{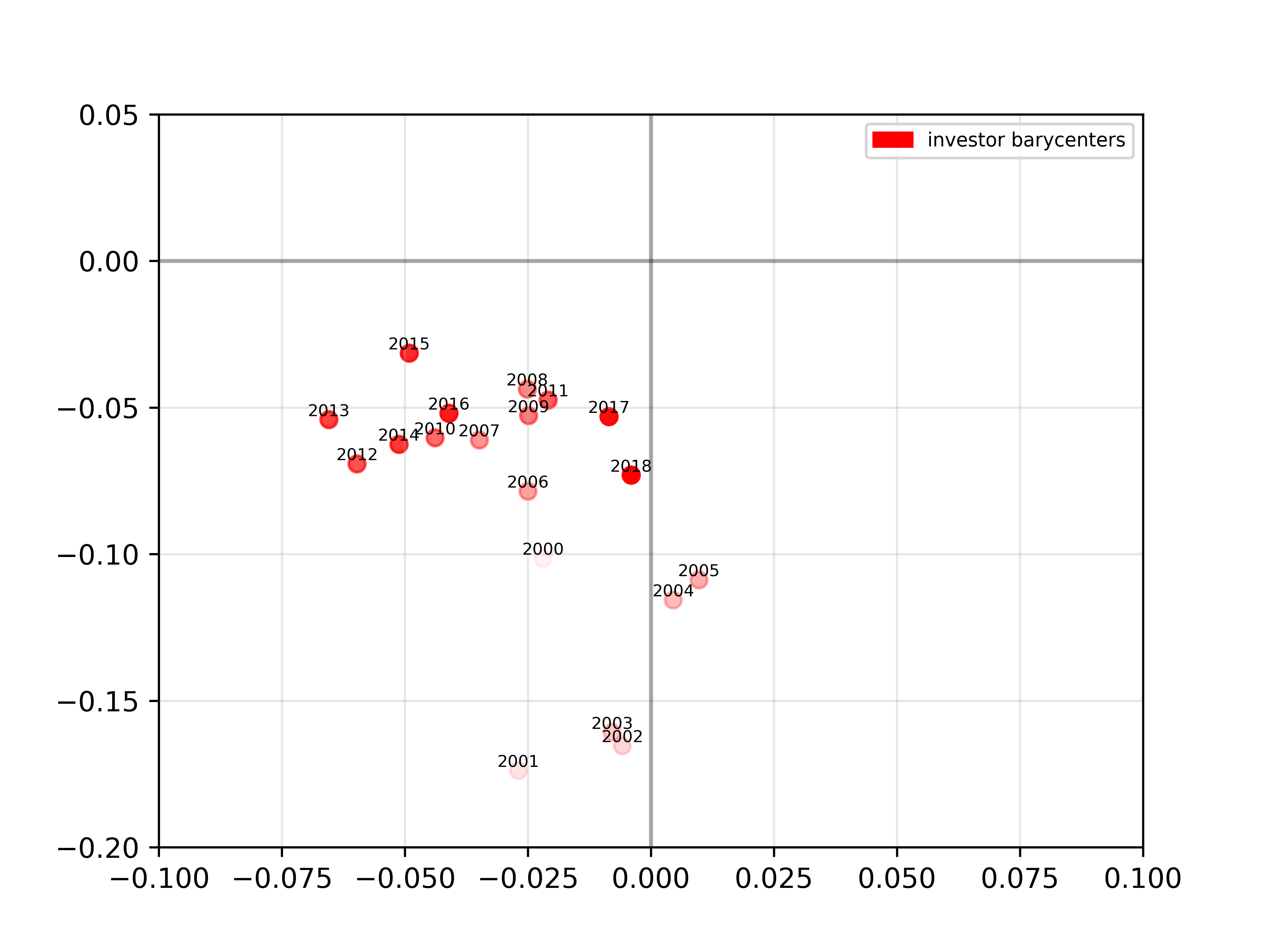}
    \includegraphics[scale = .5]{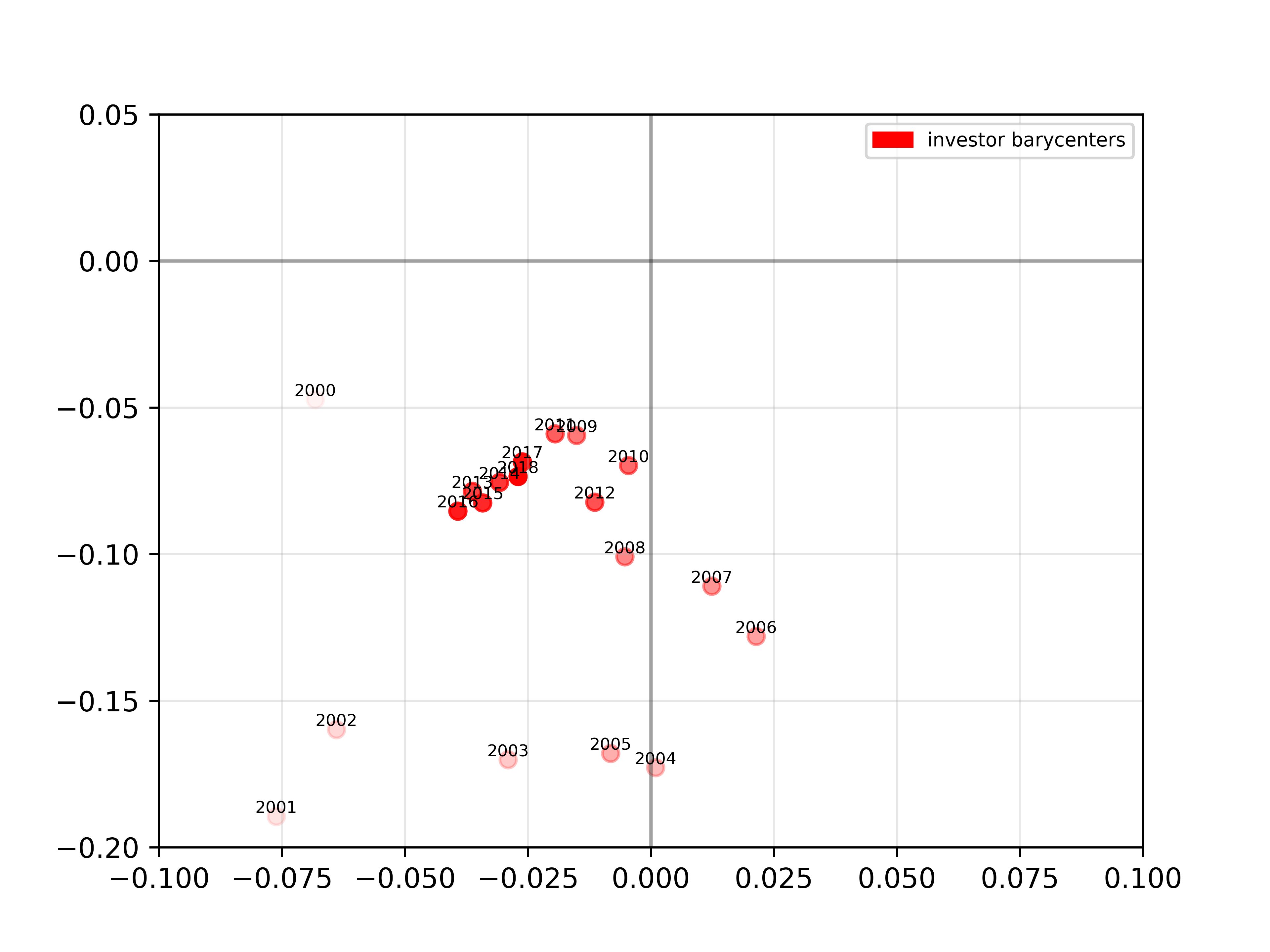}
    \caption{\footnotesize{Position of investors' barycenter at each investment stage. From top left to bottom right : Seed, Series A, Series B, Series C and later rounds aggregated.}}
    \label{fig:seedBarycenter}
\end{figure}

Furthermore, the position around which the investor community as a whole seems to "gravitate" appears to drive early stage investments when compared to later-stage ones, as evidenced by fig.~\ref{fig:seedBarycenter} where we observe a coherent grouping of early-stage investments in recent years around that center of gravity (fig.~\ref{fig:seedBarycenter}, top row).

\subsection{Temporal dynamics using TCA}
\label{subsubsec:results_TCA}
\begin{figure}[H]
    \centering
    \includegraphics[scale = .3]{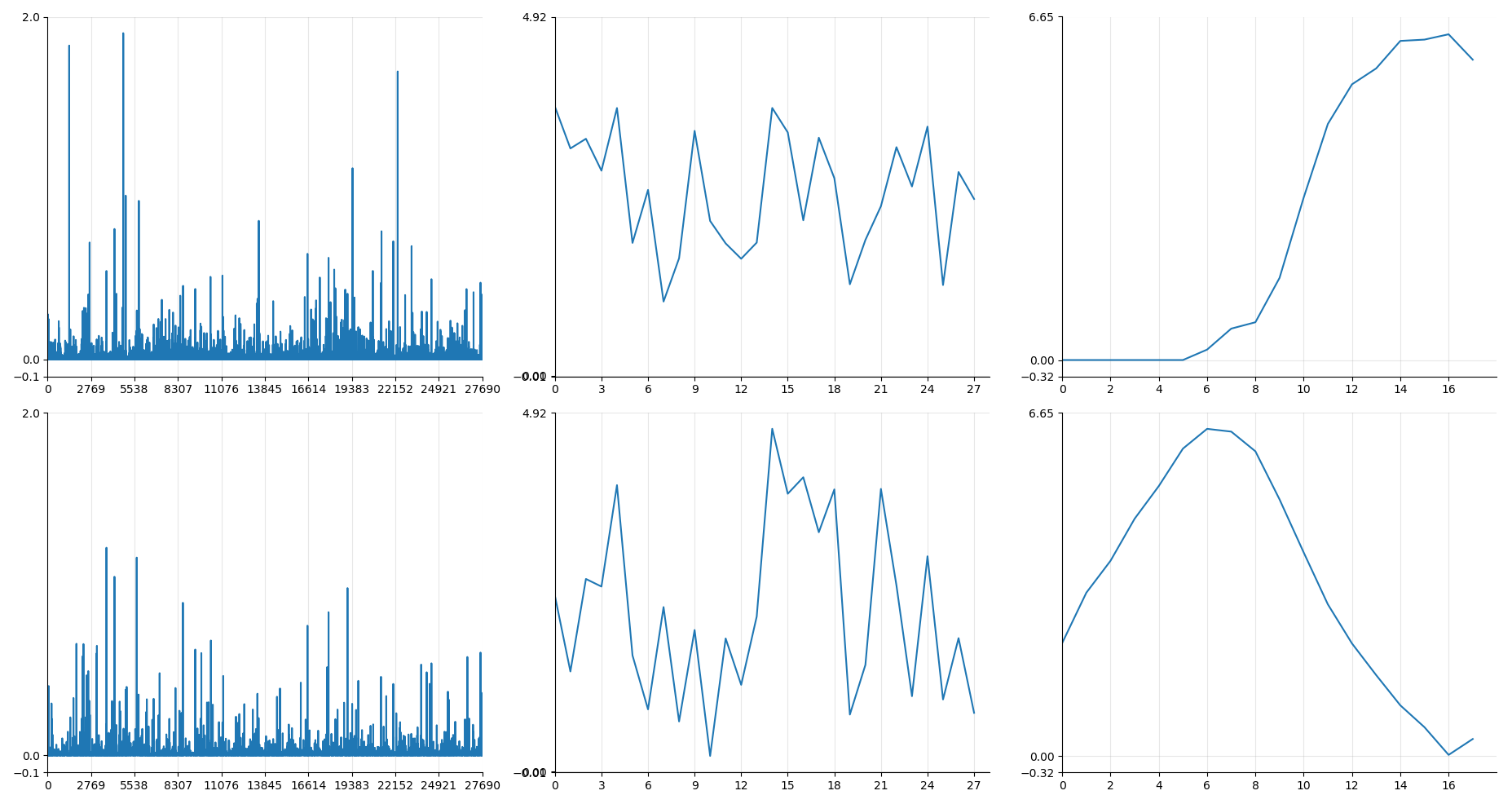}
    \caption{\small{TCA factor plots calculated between years 2000 and 2017 included for $R = 2$. The top row corresponds to the first component and the bottom row to the second component. From left to right are depicted the investor factors, the sector factors and the temporal factors.}}
    \label{fig:TCA_factors_2000-2018}
\end{figure}

The factors obtained from TCA (section \ref{subsec:TCA_method}) are presented in fig. \ref{fig:TCA_factors_2000-2018} \footnote{It should be noted that TCA removes the need to filter out the \textit{Health Care} category in investment profiles during the construction and analysis of the data tensor. Since the temporal dynamics of groups of actors are extracted separately, \textit{Health Care} investors appear as an exclusive subgroup and their impact on the dynamic of the system is limited.}. Looking at the temporal factor (third column), we observe that the first component (top line) grows in amplitude starting around 2006 while the second component (bottom line) shrinks after a maximum value in 2006. In this respect, the identity and type of the 10 individual investors with the highest \textit{investor factor} values for each component are presented in table \ref{tab:factors}. Top investors associateed with the first component are mostly accelerators, while the second component is composed of more traditional, stage-agnostic VCs, which is consistent with the fact that the first accelerators were founded around 2005-2006, kickstarting the "seed accelerator phenomenon"~\cite{Cohen2014}. Although the shape of the curves near the end of our period of study further suggests that the accelerator trend could be slowing down or changing nature, the successes of the first accelerators appear to have impacted the entrepreneurial financing ecosystem in a structural way.

\begin{centering}
\begin{table}[H]
\caption{\small{Top 10 investors and their corresponding Crunchbase classification, ranked by their first factor value.}}
\makebox[\textwidth]{%
    \begin{tabular}{|c|c||c|c|}
        \hline
        \multicolumn{2}{|c||}{\textbf{\large{1\textsuperscript{st} component}}} & \multicolumn{2}{|c|}{\textbf{\large{2\textsuperscript{nd} component}}} \\[1pt]
        \hline
        \textbf{Name} & \textbf{Type} & \textbf{Name} & \textbf{Type}\\ 
        \hline 
        Y Combinator & Accelerator & Sequoia Capital & VC (agnostic)\\ 
        500 Startups & Accelerator & Draper Fisher Jurvetson & VC (agnostic)\\ 
        Techstars & Accelerator & New Enterprise Associates & VC (agnostic)\\ 
        Rightside Capital Management & Micro VC & Intel Capital & Corporate VC (agnostic)\\ 
        SV Angel & Micro VC & Benchmark & VC (agnostic)\\ 
        Masschallenge & Accelerator & Kleiner Perkins Caufield Byers & VC (agnostic)\\ 
        Andreessen Horowitz & VC (agnostic) & Accel Partners & VC (agnostic)\\ 
        New Enterprise Associates & VC (agnostic) & Menlo Ventures & VC (agnostic)\\ 
        SOSV & Accelerator & North Bridge Venture Partners & VC (agnostic)\\ 
        First Round Capital & VC (seed stage) & U.S. Venture Partners & VC (agnostic)\\ 
        \hline
    \end{tabular}}
\label{tab:factors}
\end{table}
\end{centering}

\subsection{Investment distances and sectoral spread}
\label{subsec:distances_spread}

\subsubsection{Investment distances}
\label{subsubsec:investment_distances}

To confirm and supplement these observations, we computed the Euclidean distance, as enunciated in eq. \ref{eq:euclid}, between the yearly position of the barycenter of various selected groups and the yearly barycenter of accelerators. To do this, we revert back to the initial pre-PCA full dimensional space.
Error margins were calculated for each barycenter coordinates using the propagation of uncertainty formula.

\begin{equation}
    \label{eq:euclid}
    d = \sqrt{\sum_{i = 1}^{28}(x_{i} - y_{i})^{2}} ,
    \end{equation}
\small{where the $x_{i}$ are the coordinates of the group of interest and the $y_{i}$ are the coordinates of the barycenter of accelerators.}

\begin{figure}[ht]
    \centering
    \includegraphics[scale=.5]{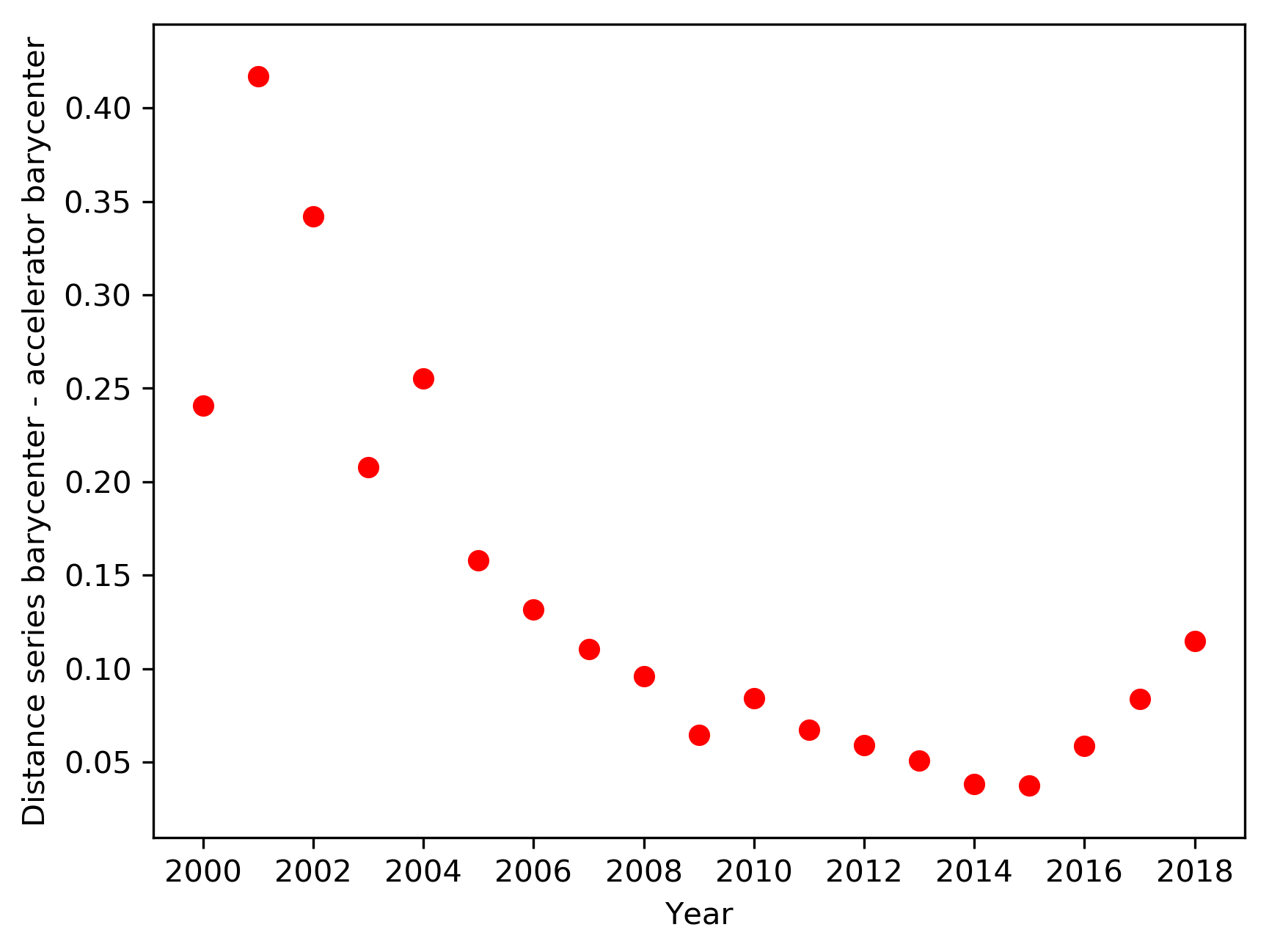}
    \includegraphics[scale=.5]{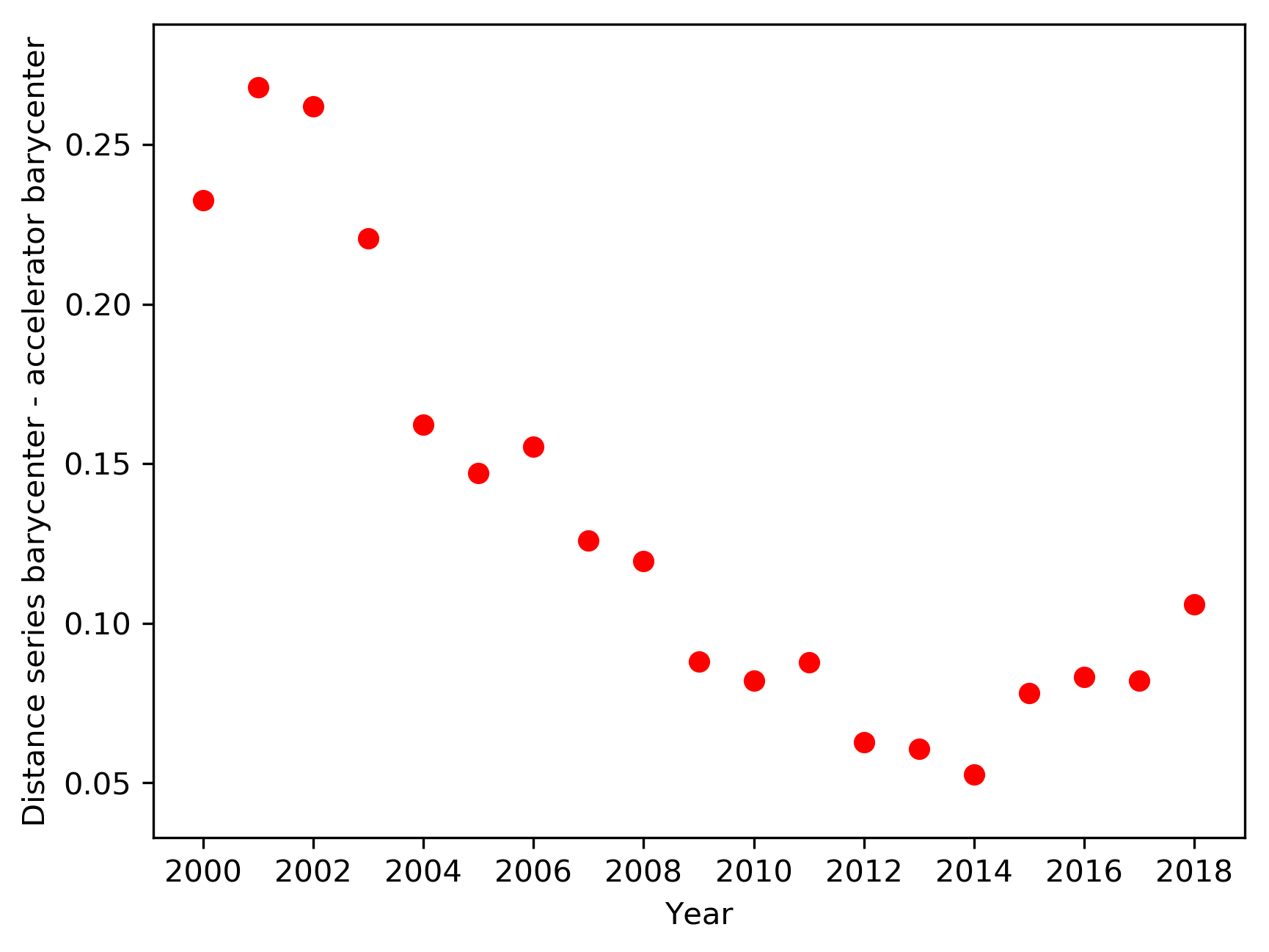}
    \includegraphics[scale=.5]{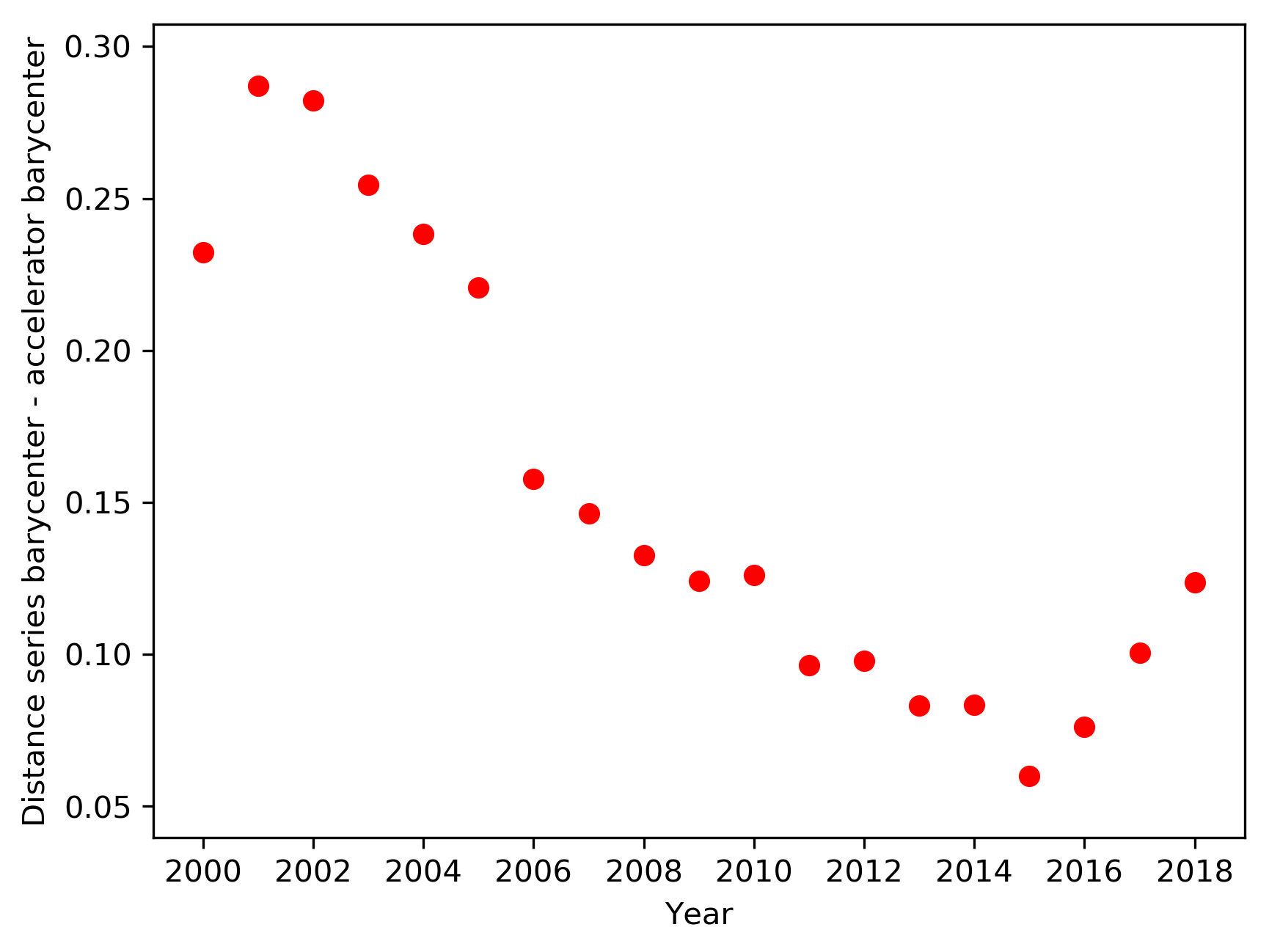}
    \includegraphics[scale=.5]{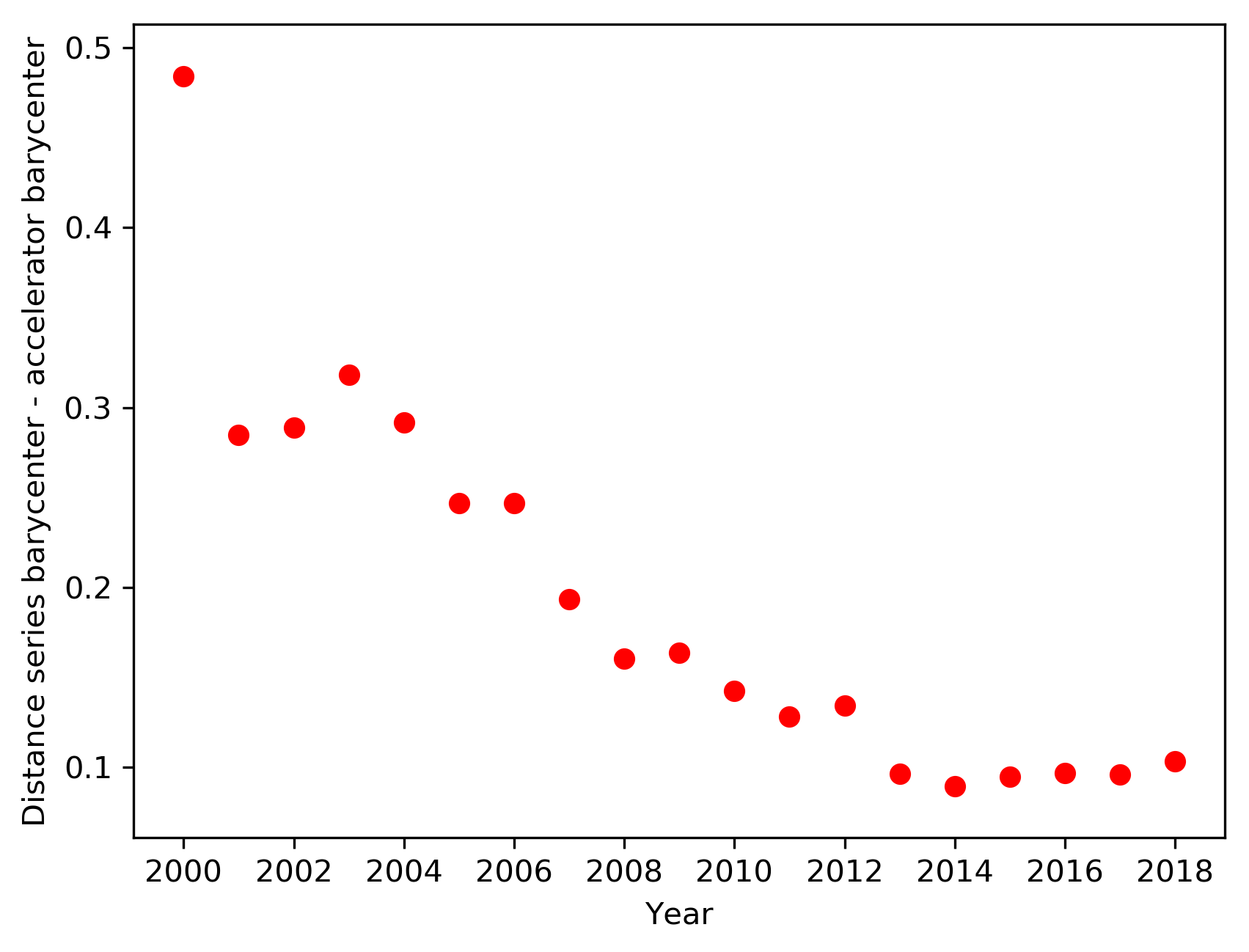}
    \caption{\small{Temporal evolution of the Euclidean distance between a barycenter for a selected series and the barycenter of the accelerators for the same year. From top left to bottom right : Seed, Series A, Series B, Series C and later.}}
    \label{fig:series2}
\end{figure}

Results are shown in figure \ref{fig:series2}. For all investment stages, the euclidean distance to the center of gravity of accelerators reduces as time goes on, reaching a minimal value around year 2014. From 2014 onwards, this distance increases again: investors, considered collectively, start moving away from the main sectors of investment of accelerators. Furthermore, the minimal value of the distance to the barycenter of accelerators increases with investment stages: the later the investment stage, the further away the barycenter of investments at this stage.

\subsubsection{Sectoral spread}
\label{subsubsec:sectoral_spread}
\begin{figure}[ht]
    \centering
    \includegraphics[scale = .9]{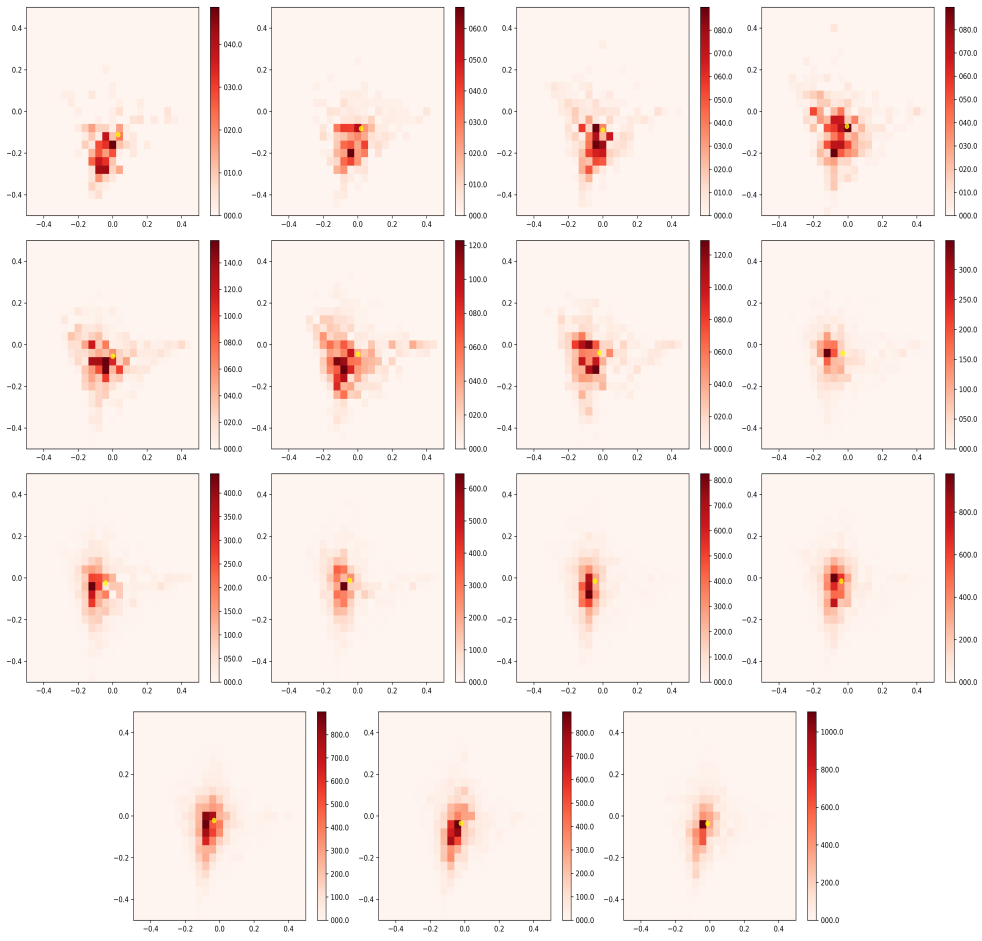}
    \caption{\small{Heatmap of the spatial distribution of investors throughout the years, from 2003 (top left) to 2017 (bottom right). The darker the cells, the more investors in that cell in sector space.}}
    \label{fig:heatmaps}
\end{figure}

\begin{figure}[H]
    \centering
    \includegraphics[scale = .4]{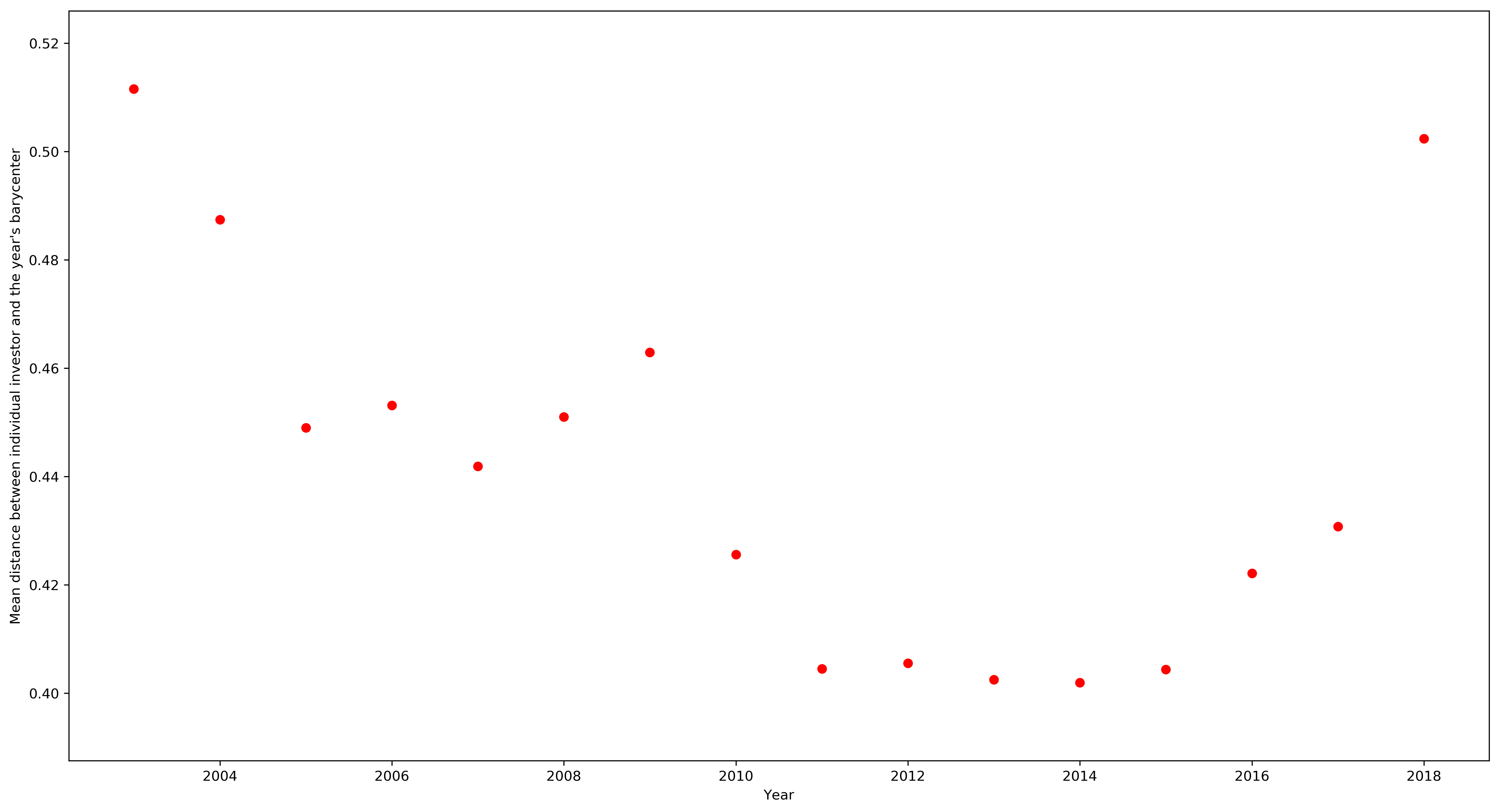}
    \caption{\small{Average Euclidean distance between individual investors and each given year's barycenter.}}
    \label{fig:average_distance_to_bar}
\end{figure}

Finally, fig.~\ref{fig:heatmaps} plots the evolution of the spatial distribution of investor portfolios in sector space between 2003 and 2017. In addition to what was previously observed, they show a marked and sudden concentration between 2010 and 2013, followed by a shift starting around 2014. These results add strikingly to the former observations, both with respect to the impact of accelerators and with respect to a move away from this kind of investment strategies in more recent years. Fig.~\ref{fig:average_distance_to_bar} plots the yearly average of the Euclidean distance between the investment strategies of all investors and the barycenter on that year, computed using the complete-dimensional sector space. Once again, it confirms that, on average, the distance between individual investors' strategies and their global yearly barycenter has decreased until 2013-2014, before noticeably drifting away starting in 2016.

\section{Discussion}
\subsection{An evolution towards lower-tech investments associated with the emergence of accelerators} 
In recent years, the barycenter of early-stage investments has moved towards a specific zone in sector space that corresponds to the zone in which accelerators were focusing their investments and that corresponds to lower-tech investment strategies. In this context, early-stage investments strategies were characterized by an increasing concentration in sector space, specially during the period 2010-2013.
This phenomenon corresponds to a collective focus on investments in more B2C and lower-tech start-ups, which were thought of as offering quicker payoffs, and which happened to be more adapted to the model of accelerators, i.e. these startups were able, contrary to many others, to quickly bring out prototypes during a brief, several months-long acceleration program.

\subsection{A recent shift away from the previous trend}
As is visible in fig.~\ref{fig:series2}, investments at all stages seem to have started to turn away from lower-tech solutions and instead to focus on more technological, now called deeptech, start-ups. This change in collective investment dynamics, from 2014-2015 onwards, corresponds also to an increased funding in the \emph{Information Technologies} and \emph{Data \& Analytics} sectors. The massive amount of artificial intelligence development that took place in recent years and the paradigm shift in this domain might be related to the change in the slope of the distance curves. This change generally signals that early- and later-stage investors are turning away from the lower-tech opportunity that they had previously been addressing. This willingness of members of the entrepreneurial ecosystem to explore new technological opportunities has the potential to lead to greater growth~\cite{Kerr2014,Timmons1986,Kortum2001}.

\section{Conclusion}
By applying common quantitative tools such as PCA and less common ones like TCA to a large dataset of venture investment rounds in startups, we were ablo to develop an original framework that allows for an in-depth study of the complex system of startup investment strategies and of its evolution over time. In this respect, the temporal dynamics of investor strategies in the US startup ecosystem since 2000 exhibit a marked evolution towards more B2C and lower-tech startups, accompanied by an increasing concentration of investment strategies, and associated with the emergence of new players in the investing ecosystem. A more recent shift away from this trend, and probably in the direction of "deeper-tech" startups, suggests that further changes are under way.
\clearpage

\end{document}